# Unzipping hBN with ultrashort mid-infrared pulses

Cecilia Y. Chen[1], Jared S. Ginsberg[2], Samuel L. Moore[3], M. Mehdi Jadidi[2], Rishi Maiti[2], Baichang Li[4], Sang Hoon Chae[4], Anjaly Rajendran[1], Gauri N. Patwardhan[2,5], Kenji Watanabe[6], Takashi Taniguchi[7], James Hone[4], D. N. Basov[3], and Alexander L. Gaeta[1,2,*]

[1]Department of Electrical Engineering, Columbia University, New York, New York 10027, USA
[2]Department of Applied Physics and Applied Mathematics, Columbia University, New York, New York 10027, USA
[3]Department of Physics, Columbia University, New York, New York 10027, USA
[4]Department of Mechanical Engineering, Columbia University, New York, New York 10027, USA
[5]School of Applied and Engineering Physics, Cornell University, Ithaca, New York 14853, USA
[6]Research Center for Functional Materials, National Institute for Materials Science, 1-1 Namiki, Tsukuba 305-0044, Japan
[7]International Center for Materials Nanoarchitectonics, National Institute for Materials Science, 1-1 Namiki, Tsukuba 305-0044, Japan
[*]a.gaeta@columbia.edu

**Manipulating the nanostructure of materials is critical for numerous applications in electronics[1,2], magnetics[3], and photonics[4,5]. However, conventional methods such as lithography and laser-writing require cleanroom facilities or leave residue. Here, we describe a new approach to create atomically sharp line defects in hexagonal boron nitride (hBN) at room temperature by direct optical phonon excitation in the mid-infrared (mid-IR). We term this phenomenon "unzipping" to describe the rapid formation and growth of a <30-nm-wide crack from a point within the laser-driven region. The formation of these features is attributed to large atomic displacements and high local bond strain from driving the crystal at a natural resonance. This process is distinguished by (i) occurring only under resonant phonon excitation, (ii) producing highly sub-wavelength features, and (iii) sensitivity to crystal orientation and pump laser polarization. Its cleanliness, directionality, and sharpness enable applications in in-situ flake cleaving and phonon-wave-coupling via free space optical excitation.**

Existing nanostructuring methods achieve nm-resolution features with cleanroom-assisted processes such as electron-beam lithography[6,7] or etching[7] or μm-resolution features with in-situ femtosecond laser-writing[8]. However, the former is time-intensive, costly, and requires multi-step processing, and the latter relies on ablation. Both leave residue or debris. Our unzipping technique generates structures on the nanoscale, orders of magnitude below the mid-IR diffraction limit, without ablation. Features are written in situ directly (resist-free) on hBN without vacuum or cryogenics in seconds, and the flake remains clean.

The in-plane hexagonal crystal structure of many 2D van der Waals (vdW) materials yields two high-symmetry axes separated by 30°, known as zigzag and armchair. In hBN, the TO($E_{1u}$) phonon at 7.3 μm corresponds to in-plane atomic motion parallel to the zigzag axis (Figure 1b inset) where boron and nitrogen are displaced in opposite directions. This optical phonon is dipole-active and lies in the laser-accessible mid-IR regime due to the material's light constituent atoms. Coherent resonant excitation of the phonon at pulse intensities of 10 TW/cm$^2$—within the linear phonon-driving regime and far below the estimated laser-induced damage threshold of 50 TW/cm$^2$—was calculated to yield transient strains of 5% of the equilibrium lattice constant[9]. Furthermore, nonlinear phononics has been demonstrated to dynamically modify the symmetry of materials[3,10–13], resulting in structural phase-transitions and ferroic behavior. Preferential orientation of flake fracture and crack formation and propagation has been studied previously in hexagonal vdW materials subject to exfoliation forces[14]. Predictably, the histograms cluster around the armchair and zigzag axes, with an angular spread about each orientation. Graphene and hBN have roughly equal preference for armchair and zigzag, while 2H-MoS$_2$ and analogous transition-metal dichalcogenides (TMDs) display strong directional preference[14,15].

In this study, we take a novel approach to deterministically inducing flake fracture by exploiting the natural vibration of atoms and driving them coherently with mid-IR pulses. Here, pulse intensities of 50–65 TW/cm$^2$ generate larger atomic displacements than previously studied and lead to macroscopic material structuring in the form of controllable, localized, rapid crack propagation. These displacements amplify lattice-scale strain and prompt organized fracture along an imposed symmetry. This approach allows us to introduce atomic-scale line defects to ultra-high-purity exfoliated hBN flakes by directly accessing the TO($E_{1u}$) mode from free space, which we demonstrate in samples 24 to 76 nm in height. The features are seeded at a random point within the laser-driven spot, possibly originating from an intrinsic defect, and unzip from the tip with largely unidirectional growth. The speed of formation is estimated to be on the order of 100 μm/s. To the best of our knowledge, this work represents the first demonstration of optically generated single line defects using a wavelength far below the material bandgap, and this unzipping phenomenon occurs only through resonant driving of hBN.

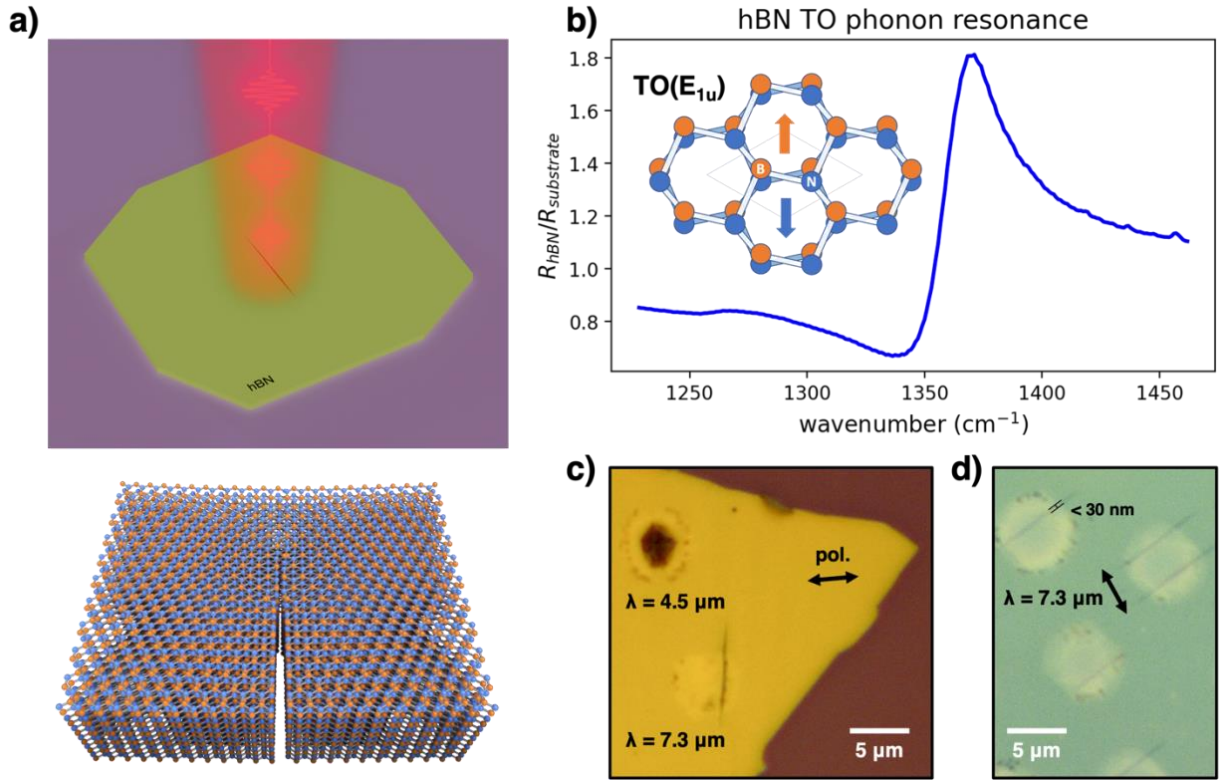

**Figure 1. A phonon-resonant effect** | "Unzipping" occurs only when hBN is strongly driven at its TO phonon resonance and yields ablation-free line defects. **(a)** (top) In this experiment, a pulsed mid-IR laser is focused onto an hBN flake, producing a localized edge or "zip." (bottom) The zip is oriented along the armchair axis. **(b)** Fourier-transform infrared spectroscopy (FTIR) linear reflectance spectrum about the hBN TO($E_{1u}$) phonon resonance on pristine hBN relative to the Si/SiO$_2$ substrate, centered at 7.3 μm (1367 cm$^{-1}$). (inset) AA'-stacked hBN illustrating the mode of interest. We directly drive this mode with a laser tuned to 7.3 μm to subject the crystal to high in-plane lattice-scale strain. **(c)** Comparison of off- and on-resonant ultrafast irradiation of a 70-nm flake at 4.5 and 7.3 μm. The latter results in a highly sub-wavelength line defect (zip); burning is absent, and the line is oriented roughly perpendicular to the laser polarization. Off-resonant irradiation at 4.5 μm generates a wavelength-scale burned spot and lacks polarization dependence. **(d)** A series of clean, parallel zips produced by a perpendicularly polarized laser on a single flake 38 nm in height. The width here measures <30 nm (Supplementary Figure 6).

Supplementary Figure 1 illustrates the experimental setup in which we gate the high-intensity pulses using a manual shutter until the damage threshold is reached. In Figure 1c, a single flake was irradiated at neighboring spots by a tunable femtosecond laser, both at (7.3 μm) and away from (4.5 μm) its optical phonon resonance. The polarization was identical in both cases, and irradiation continued until the first indication of structural change to the flake. We observe several critical differences. A line defect appears at the on-resonant spot, while the off-resonant spot is burned and ablated. The minimum achievable zip width is measured to be <30 nm despite being generated by a 7.3 μm laser, making the features highly sub-wavelength in scale. With off-resonant irradiation, the size of the ablated spot is on the order of the irradiation wavelength, as expected for conventional laser damage. Line defects do not appear at the off-resonance damage threshold even with the same polarization and flake orientation, further confirming that unzipping is associated with on-resonance driving. The absence of ablation distinguishes unzipping as a method of gentle and wavelength-selective defect creation which also displays spatial photoluminescence (PL) localization (Supplementary Figure 2).

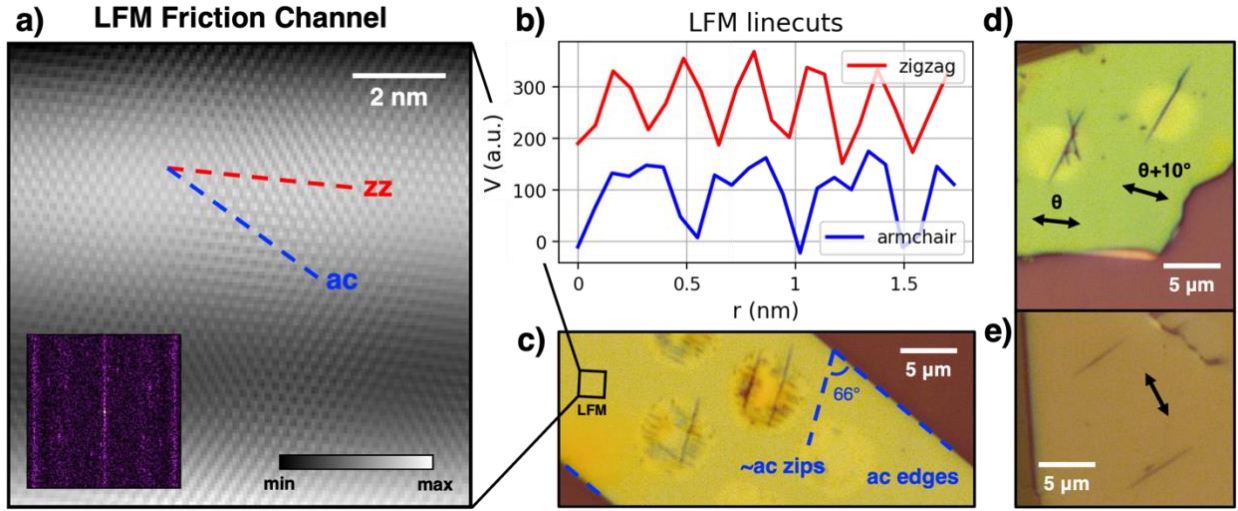

**Figure 2. Polarization dependence** | Unzipping is sensitive to the pump laser polarization. **(a-c)** LFM determines the crystal orientation of a bulk hBN flake and hence the unzipping direction. The scan region is marked in (c). **(a)** LFM friction channel image after filtering, with zigzag and armchair directions marked. (inset) 2D-FFT of unfiltered friction channel. **(b)** Linecuts along the zigzag and armchair directions yield periodicities of 29 and 50 Å, respectively, confirming the measurement. The y-axes are offset for clarity. **(c)** The 70-nm-thick flake imaged with LFM. The zips measure 66° from the armchair-oriented flake edges, making them nearly parallel to an armchair axis of the crystal. **(d-e)** The unzipping phenomenon occurs independently of the choice of substrate. **(d)** Unzipped lines on a 60-nm-thick hBN flake on $SiO_2$/Si. The sensitivity of unzipping to pump laser polarization is evident in the transition from an X-shaped line defect to a parallel single line under a 10° shift in polarization. **(e)** Unzipped lines on an 84-nm-thick hBN sample on sapphire. Zips generated by the same polarization are parallel.

Since unzipping is the product of selective phonon driving, it inherits symmetries from the crystal. As expected for a system with rotational symmetry, the ease of unzipping hBN depends on the relative angle between the laser polarization and the crystal axes. Unzipped lines form roughly perpendicular to the polarization, with a deviation of ±15°. For a given flake orientation, not every polarization can produce a line defect. A qualitative 6-fold symmetry in "unzippability" was revealed upon sweeping all relative angles between the sample orientation and pump polarization like in Supplementary Figure 3. The distribution of these polarizations is periodic: the sector width of polarizations that unzip hBN is ≤50° and repeats every 60°. Polarizations falling outside of these ranges will produce a dense crosshatch pattern (Supplementary Figure 3-D). Nevertheless, in contrast to the off-resonant example, the structural changes retain a straight-edged, ordered geometry. Furthermore, the optimal polarization within each sector yielding the highest-quality unzipped lines that are atomically sharp, clean, and straight is parallel to the zigzag axis, or perpendicular to the resulting zip. The farther the polarization strays from optimum, the more resistant the flake is to unzipping (requiring prolonged laser exposure and exhibiting slower crack propagation) and the more likely the lines will be wider, kinked (Figures 2d, 3a, 4d), feathered (Figure 2d), or produce light debris (Figure 1c, Supplementary Figure 3-A).

Regardless of zip quality, all line defects created via phonon-resonant excitation are sharply tapered at the ends, which supports the picture of unzipping as originating from bond rupture. Driving the TO($E_{1u}$) phonon stretches bonds parallel to the zigzag axis, which we expect leads to fracture primarily in the direction perpendicular to the applied strain, or along the armchair axis. To test this hypothesis, we performed lattice-scale lateral force microscopy (LFM)[16,17] on our bulk exfoliated flakes to determine the crystal orientation, since the standard second-harmonic generation (SHG) method is unsuitable in hBN beyond the few-layer limit[18–20]. Figure 2a displays the filtered friction channel image with its raw 2D-FFT. Figure 2b shows the friction channel linecuts along the zigzag and armchair directions. Periodicities were measured to be $T_{zz}$=2.9 Å and $T_{ac}$=5.0 Å, respectively. The ratio $T_{ac}/T_{zz}$=√3 as expected for a hexagonal lattice, despite a proportional deviation from the accepted hBN lattice periodicities of 2.5 and 4.3 Å[21]. From the LFM measurements, we conclude that the flake in Figure 2c unzips approximately along the armchair direction (66° from the armchair flake edge) even when created under suboptimal conditions.

In addition to single unzipped lines, there are variations on unzipping that can appear under comparable irradiation conditions. As seen in Figures 1d and 2c, two independent yet parallel lines may form simultaneously. Figure 2d displays an example of the rarer X-shaped unzipping due to its high sensitivity to relative polarization. Two kinked, near-armchair zips separated by ~50° result from symmetric activation. While the individual zips are no longer perpendicular to the driving polarization as in the standard case, the angle bisector is, and shifting the polarization by just 10° recovers a single zip parallel to one of them. Similar atomic-scale line defects have been generated in monolayer graphene using e-beam irradiation[15] (although exhibiting a slower crack propagation speed of 1 μm/s) but without polarization dependence or X-shaped line defects, again affirming the role of laser-phonon driving in our experiment.

An added feature of the unzipping technique is that it is consistent and robust. For a given flake at fixed orientation driven by the same allowed pump polarization, every zip formed will be parallel. This is illustrated in Figures 1d and 2e. Furthermore, line defects can be generated even with a moderately misaligned laser, slightly irregular beam shape, variation in average pump intensity and pulse-gating pattern, and/or suboptimal pump polarization, and across flake thicknesses varying by tens of nanometers. When a flake properly unzips, it is clean and does not exhibit ablation damage. The flake in Figure 2c unzipped along the armchair axis despite poorer beam shape and alignment, which resulted in a zip that was not as sharp, clean, and straight. (The irregular streaking within these irradiated spots is not unzipping, and the flake appears smooth under AFM. See Supplementary Figure 4b.)

Finally, the unzipping effect itself occurs independently of the substrate. Features were produced primarily on flakes on SiO₂/Si, but unzipped lines also form on hBN on sapphire (Figure 2e). However, unzipping flakes of similar thickness on sapphire is more difficult due to poorer flake adhesion, resulting in greater sensitivity to the irradiation parameters. These flakes tend to unzip and immediately rip or peel from the substrate. To combat the less robust nature of unzipping on sapphire, the average pump intensity was reduced by a factor of 3, and we operated within a narrower viable fluence window.

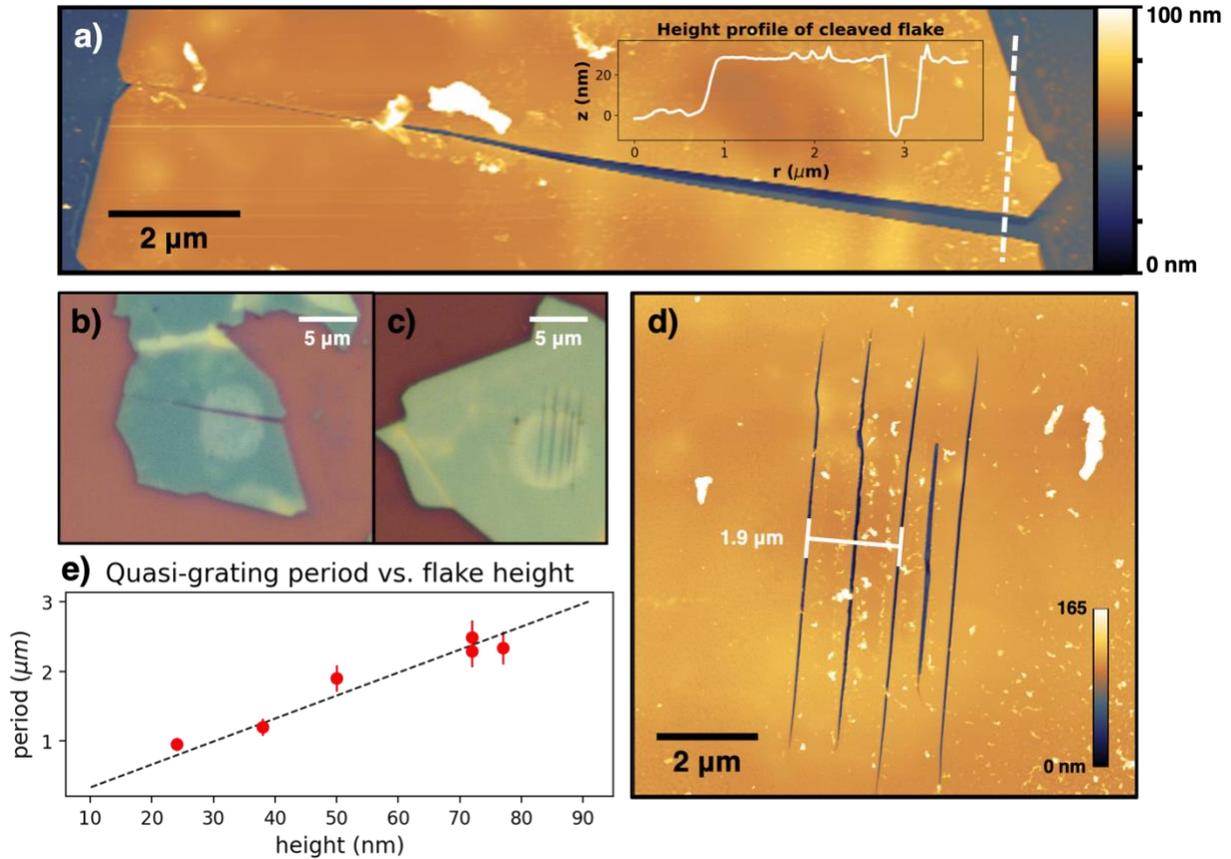

**Figure 3. Applications in flake patterning and nanostructuring** | All-optical in-situ patterning of hBN flakes using the mid-IR phonon-resonant technique. **(a-b)** Unzipping can be controllably and boundlessly extended once initiated, achieving an ultra-high aspect ratio. **(a)** AFM topography: A 24-nm-thick flake is cleaved in two by extending an initial unzipped line in both directions. A slight edge offset at the left and right ends of the cleavage line reveal that the bottom section has rotated and shifted along the substrate. (inset) Height profile along the dotted linecut confirms full separation of the cleaved sections. **(b)** Micrograph: The initial unzipped line is localized within the slight discolored spot. **(c-d)** Quasi-periodic gratings generated by the unzipping technique on a 50-nm-thick flake, represented in a micrograph (c) and topographic AFM image (d). **(e)** Linear trend displayed by grating period versus flake height.

One application of unzipping is an all-optical, orientation-selective, in-situ method for cleaving or patterning flakes. In Figures 3a-b, unzipping cuts completely through thinner <30-nm-thick flakes.

Once a zip is created (localized within the faint circle), it can be elongated boundlessly with ease by shifting the pump beam incrementally along the unzipping axis with a piezoelectric stage. Even without a stage, the line can be extended cleanly by irradiating slightly beyond its existing endpoints. This technique may be used to cleave a single hBN flake for self-oriented stacking in a moiré homostructure[22]. Similarly, adjacent regions can be irradiated with different polarizations, and the resulting zips will snap together at 60° angles (Supplementary Figure 5), opening the door to user-defined flake shapes and line defect designs. These examples illustrate hBN's potential as a patterned, orientation-selective layer, either by itself, as a mask[7], or in a 2D heterostructure, especially since it is mechanically tough[23], has a large bandgap[24], and is widely used as an encapsulating material[25].

The unzipping mechanism can also generate quasi-periodic gratings (Figures 3c-e) at slightly higher irradiation fluences. These gratings emanate from an initial unzipped line, mainly in one direction, at a speed of 10 μm/s. In some cases, they can be introduced separately by irradiating an existing single zip. Grating period follows a linear trend with flake height (Figure 3e). We hypothesize that these gratings form when phonon-polaritons[26] are launched from the initial zip and rip up the crystal as they propagate, although an investigation into the exact mechanism is beyond the scope of this work. Notably, unzipping and their quasi-periodic gratings are fundamentally distinct from nanogratings[27] and laser-induced periodic surface structures (LIPSS)[28]. The nanostructures described here display atomically sharp, tapered lines with a low duty cycle, are associated with a resonant phonon effect instead of plasmas, and are oriented along the armchair crystal axes. Quasi-periodic gratings may find applications in enhancing wave coupling from free space or as a standalone nanopatterning technique.

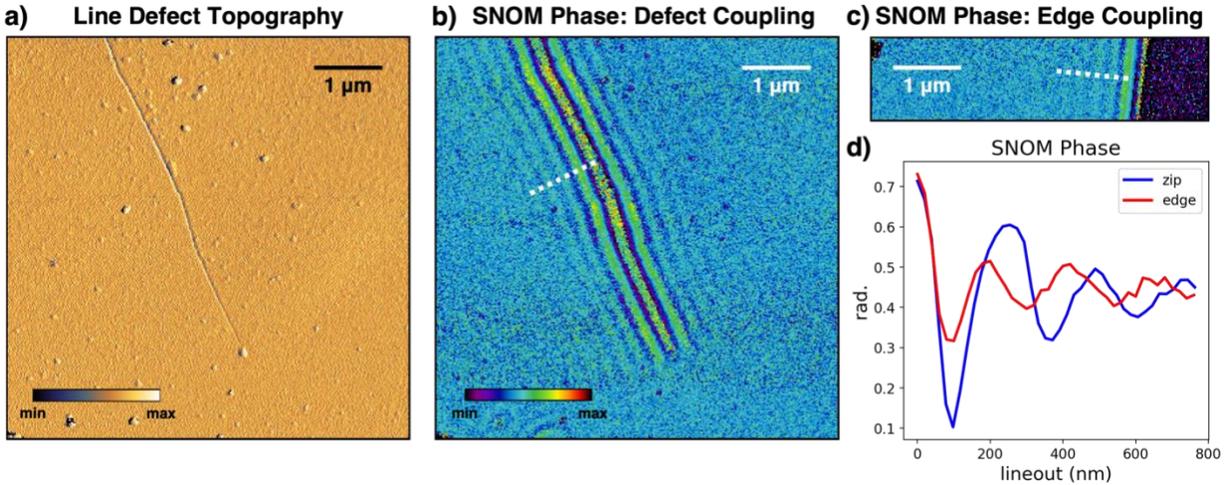

**Figure 4. Atomically sharp edges** | Unzipped lines are atomically sharp and exhibit highly efficient coupling to phonon-polaritons. This flake is 38 nm in height. **(a)** AFM topography of a kinked (suboptimal) zip. **(b-c)** Near-field phase images under 1494.5 cm$^{-1}$ excitation, as probed by s-SNOM. **(b)** Phonon-polariton coupling via an unzipped line. **(c)** Coupling via a natural flake edge formed by mechanical exfoliation. **(d)** Unzipped lines are atomically sharp and can outperform flake edges in coupling to phonon-polaritons, as evidenced by a comparison of modulation depth and decay rate along the dotted linecuts in (b) and (c).

Figure 4 demonstrates efficient coupling from free-space optical excitation to polaritons in hBN via unzipped nanostructures. Figure 4d compares linecuts of scattering-type scanning near-field optical microscopy (s-SNOM) phase images from excitation within the hBN upper Reststrahlen band: phonon-polaritons launched by the unzipped line exhibit much greater modulation depth and slower decay than those by a natural edge on the same flake. We conclude that unzipped "artificial edges" can outperform exfoliated edges in coupling to polaritons and give the user freedom of edge placement. Compared to conventional coupling methods[4], unzipping yields a directional phase front, as opposed to launching from an AFM tip[26,29], and is a quick, cleanroom-free technique, unlike e-beam patterning[4] or depositing gold nanostructures[4,30].

In summary, strong resonant driving of hBN at its 7.3 μm TO($E_{1u}$) phonon resonance allows us to unzip the flake, or place atomically sharp edges within the flake interior parallel to the armchair crystal axes. In contrast to the flake-burning typical of strong off-resonant irradiation, this gentle and debris-free technique of selectively driving a vibrational mode to generate localized, directional bond strain within the crystal is the first demonstration of optically induced strain strong enough to cause flake fracture. We show that the sharp edges generated in situ by unzipping hBN exhibit highly efficient coupling to phonon-polaritons, can cleave flakes, and localize PL emission and envision future applications in custom flake shaping, mask patterning, and oriented hetero-/homo-structure stacking.

## Methods

*Sample preparation*

Ultra-high-quality AA'-stacked hBN (intrinsic defect density $10^9$ cm$^{-2}$) is mechanically exfoliated using low-residue Scotch Magic Greener tape onto 285 nm SiO$_2$-on-Si and sapphire substrates which were first subject to an O$_2$ plasma treatment to remove adsorbates. No annealing was performed after exfoliation.

*Phonon-resonant irradiation of hBN*

The hBN flakes are irradiated with a 1 kHz pulsed mid-IR laser tuned to the hBN TO($E_{1u}$) phonon at 1367 cm$^{-1}$ (7.3 µm). We achieve this output with a laser system consisting of a Ti:sapphire mode-locked oscillator (KMLabs Griffin), Ti:sapphire chirped pulse amplifier with regenerative amplification outputting 6 mJ pulse energy at 1 kHz repetition rate (Coherent Legend Elite), optical parametric amplifier (Light Conversion HE-TOPAS Prime), and a subsequent difference-frequency generation module (Light Conversion NDFG). The resulting mid-IR pulses measure 120 fs in pulse width and 1.5 µm FWHM.

The beam is routed to the sample in reflection geometry at normal incidence with a mid-IR dichroic mirror (ISP Optics BSP-DI-25-3) and focused with a 40x reflective objective (Thorlabs LMM-40X-P01, 0.5 NA). The incident average power is 250 µW. Pulse selection is performed manually with an electronic shutter (Melles Griot) as we monitor the formation of unzipped lines on a CCD camera mounted in the reflection path. The total laser fluence required to achieve an initial unzipped line defect is on the order of $10^3$ J/cm$^2$ (variable, depending on relative polarization and local flake thickness and uniformity), adjusted with a combination of ZnSe reflective ND filters (Thorlabs), mid-IR polarization optics, pulse gating settings (shutter speed and repetitions), and an optional band pass filter at 7500 ± 50 nm (Thorlabs). hBN on SiO$_2$/Si unzipped with an average pulse intensity of 50 TW/cm$^2$; flakes on sapphire required lower peak power due to weak substrate adhesion. All defect creation was performed under ambient conditions.

The angle between the laser polarization and crystal orientation was controlled by either rotating the sample relative to a fixed linear polarization or adjusting the pump polarization using mid-IR polarization optics (Alphalas tunable zero-order waveplates, Thorlabs ZnSe wire grid polarizers), which yielded equivalent results. Extending the unzipped line requires significant pump power reduction which we achieve with the 7.5 µm band pass filter (Thorlabs). It cuts the total power by 85% while still maintaining a spectrum that overlaps the TO phonon resonance in Figure 1b. This can also be substituted for additional ND filters. The most uniform line defect extension is achieved by incrementally translating the sample on a piezoelectric stage. Another technique is to irradiate an adjoining spot—the independently-generated unzipped lines will merge.

*Off-resonant irradiation of hBN*

To confirm that unzipping is a resonant effect, we repeat the defect generation process at an off-resonant mid-IR wavelength. The experimental setup is identical, with the output of the NDFG module now tuned to 4.5 μm ($\tau_p$=100 fs, 1 μm FWHM). The fluence required to straddle the damage threshold here is very similar to that of the resonant case (at minimum, the same order of magnitude).

*AFM topographic imaging*

AFM topography was imaged on a Bruker Dimension Icon in automated tapping (ScanAsyst) mode with ScanAsyst-Air probes at 0.5 Hz scan rate and 128-pixel resolution. Plane leveling and post-processing was done in Gwyddion[31].

*Lattice-scale imaging of hBN crystal orientation*

We determine the crystal orientation of hBN flakes using lateral force microscopy (LFM). Friction channel scans were taken on a Bruker Dimension Icon with a silicon nitride tip on a silicon nitride cantilever (Bruker DNP-C, nominal k=0.24 N/m) at 2.5 Hz scan rate in constant height mode. Residual noise from the XY sensors can affect scans at the angstrom level, so the XY closed loop control parameter should be turned off. A fresh anti-static ionizing cartridge containing alpha particle-emitting polonium-210 (StaticMaster CPSMR3) was placed near the sample to neutralize electrostatic effects impeding tip engagement and imaging. We then performed plane leveling and 2D-FFT filtering on the raw images in Gwyddion.

*Scattering-type scanning near-field optical microscopy (s-SNOM)*

Near-field images were obtained on a Neaspec GmbH neaSCOPE system equipped with a PtIr-coated AFM tip (ARROW-EFM, 20 nm tip radius) operating at 285 kHz tapping frequency. We use a broadly tunable light source (Daylight Solutions MIRcat quantum cascade laser) chosen to coincide in frequency with the hBN upper Reststrahlen band[26] (1368–1610 cm$^{-1}$); the signal scattered off the probe tip is sent to a mercury-cadmium-telluride (MCT) detector. We interferometrically detect and demodulate the signal at harmonics of the tapping frequency via the interferometric pseudo-heterodyne technique[32] to obtain background-free near-field signals with amplitude and phase information. The resulting images demonstrating phonon-polariton propagation in the vicinity of the hBN zip and a natural edge were then processed in Gwyddion. Topography information was obtained concurrently.

*Fourier-transform infrared spectroscopy (FTIR) and photoluminescence (PL) spectra*

The far-field FTIR reflectance spectrum of pristine hBN was recorded with a Bruker Vertex 80v FTIR spectrometer coupled to a Bruker Hyperion II microscope. The aperture of the mid-IR illumination lamp was set to 25 μm. Spectra were acquired with 1000 averages and 2 cm$^{-1}$ resolution. The $SiO_2$/Si substrate was used as a reference.

PL spectra from 532 nm excitation were acquired on a Bruker Senterra I with a 50x objective lens (Olympus MPlan N, 0.75 NA) and 2 mW of laser power. The signal was integrated over 15 seconds.

*Scanning electron microscopy (SEM)*

SEM imaging was carried out using a Zeiss Sigma VP SEM. To avoid an electron charging effect, the accelerating voltage was set to 1 kV.

**Data Availability**

The data generated and/or analyzed during this study are available from the corresponding author upon reasonable request.

**Acknowledgments**

We thank M. Lipson and P. J. Schuck for helpful discussions and advice. C.Y.C. thanks John Thornton from Bruker Corporation for his support and expert guidance on performing LFM measurements. The work was carried out, in part, in the Shared Materials Characterization Laboratory of the Columbia Nano Initiative (CNI) Shared Lab Facilities at Columbia University.

This work is supported as part of Programmable Quantum Materials, an Energy Frontier Research Center funded by the U.S. Department of Energy (DOE), Office of Science, Basic Energy Sciences (BES), under award DE-SC0019443. C.Y.C. acknowledges support from the NSF Graduate Research Fellowship Program DGE 16-44869. K.W. and T.T. acknowledge support from the JSPS KAKENHI (Grant Numbers 19H05790, 20H00354, and 21H05233).

The authors declare no competing interests.

# Supplementary Material

*Phonon-resonant irradiation setup for unzipping hBN*

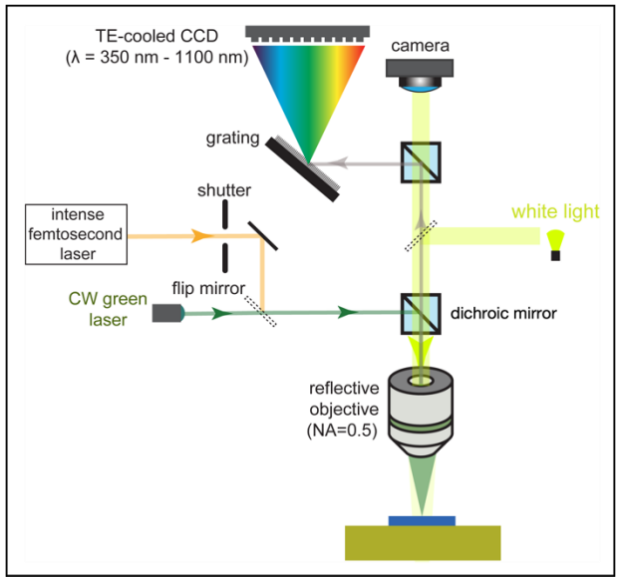

**Supplementary Figure 1.** Experimental setup for mid-IR excitation (in dotted gray outline) and preliminary PL measurements.

*Localization of PL emission*

hBN is a promising medium for hosting stable, bright, room-temperature single photon emitters[1], and unzipping may present an avenue for locally and controllably writing these quantum defects without laser ablation[2,3] or e-beam writing[3,4]. As seen in the room-temperature PL spectra below (Supplementary Figure 2), defect emission from unzipped lines takes the shape of a broad hump ranging from 600 to 800 nm with prominence around 650 nm. This coincides with the natural distribution of individual emitters and spectra of emitter ensembles in hBN[5,6]. However, the exact nature of the defects is presently unknown. We suspect that a large collection of atomic defects and dangling bonds are localized within the unzipped or irradiated region, and that low-temperature[1,7] coupled with polarization-resolved[8] measurements may reduce emitter linewidths and density enough to resolve defects spectrally and reveal distinct zero-phonon lines.

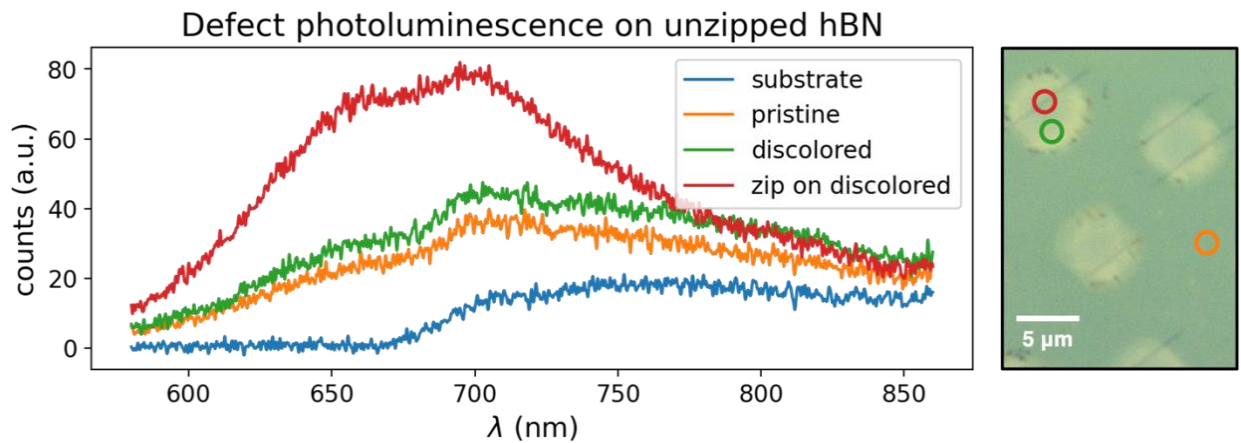

**Supplementary Figure 2.** PL emission from an unzipped region of hBN under continuous-wave 532 nm excitation, compared to the Si/SiO$_2$ substrate, a pristine region of the same flake, and an adjacent, lightly discolored region indicating moderate laser exposure below the damage threshold. Corresponding measurement locations are marked on the micrograph.

*Proper versus failed unzipping*

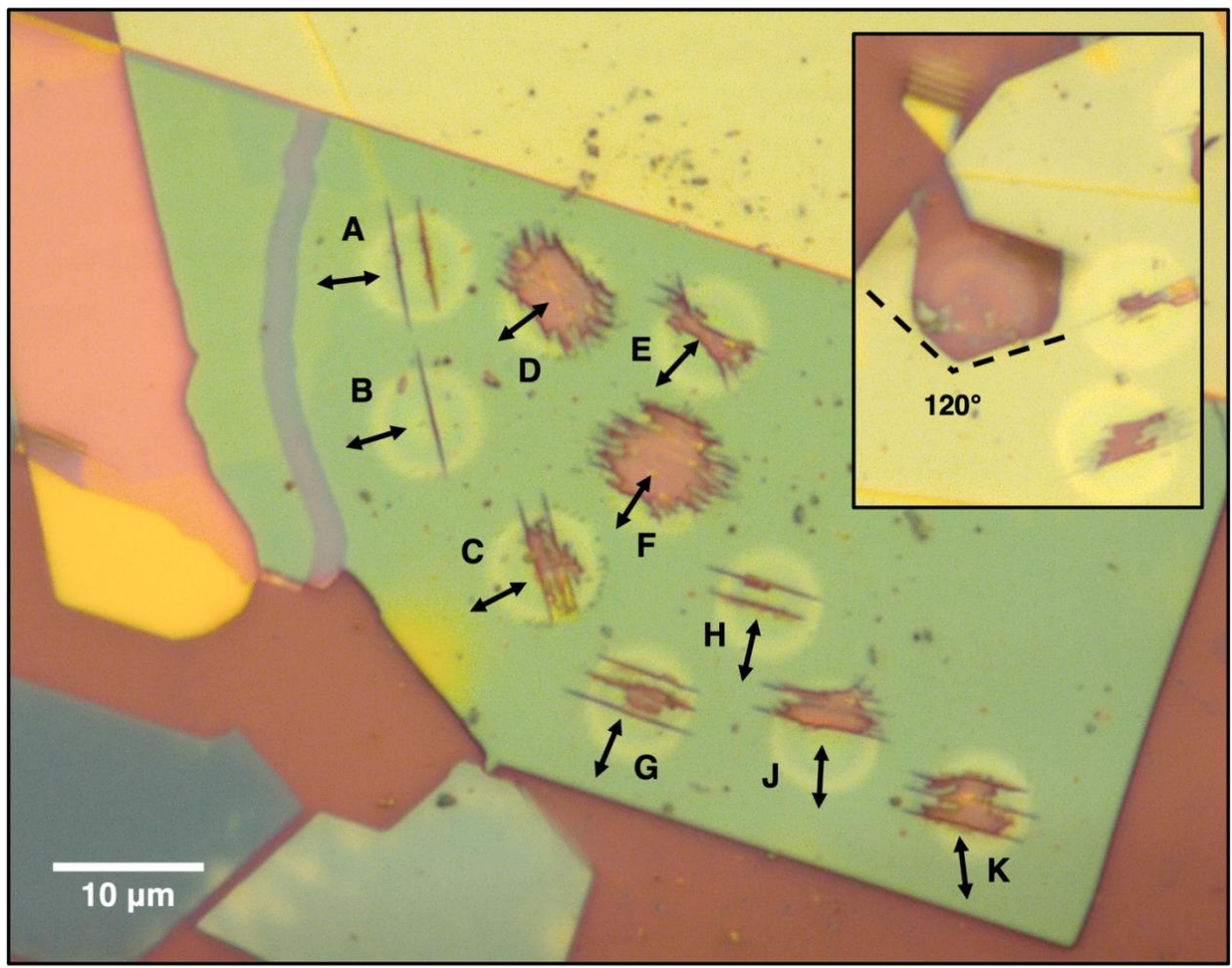

**Supplementary Figure 3.** A comparison of proper vs. failed unzipping due to different relative pump polarizations and fluences. The polarization is rotated through 90° in increments of 10° across spots **A-K**. The debris is from failed and suboptimal zips; proper zips are debris-free. Spots **A-B** are considered proper zips formed under the correct laser polarization relative to the crystal symmetries and within the acceptable fluence window; spot **C**, with lines along the same unzipping axis, was subject to both a less optimal polarization and a slightly higher fluence. Spots **G-K** unzipped (sub-optimally, due to excessive fluence) along an adjacent unzipping axis 60° away from the previous one. Crosshatched patterns in spots **D-F** result from polarizations that do not support unzipping. However, the constituent lines within the crosshatches are roughly parallel to the flake's preferred unzipping directions. **(inset)** Resonant irradiation in pursuit of unzipping (at any pump polarization) sometimes unpredictably ejects a polygonal piece of the flake, exposing the substrate. The resulting void possesses edges that follow the flake's unzipping directions (visible to the right), and at least one of its angles measures ~120°. Although the void in this example is located near a flake edge, it can occur just as readily in the flake interior. We hypothesize that the unpredictability is due to local flake conditions on the macro- and nano-scale, such as delamination or a higher concentration of intrinsic defects.

*Lateral force microscopy (LFM)*

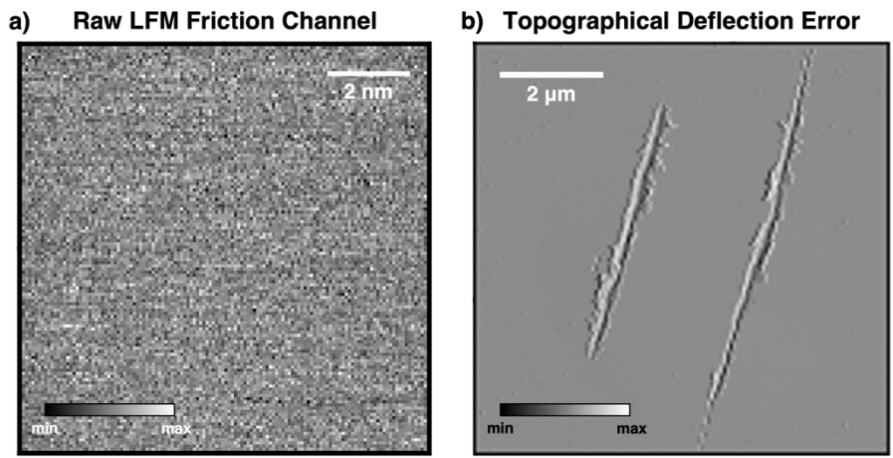

**Supplementary Figure 4.** **(a)** Raw friction channel scan corresponding to Figures 2a-c. **(b)** Topographical deflection error image corresponding to the double-unzipped spot in Figure 2c. The flake surface remains smooth, even after unzipping with suboptimal irradiation conditions.

*From lines to shapes: Connecting zips at 60° angles*

We can exploit the sensitivity of unzipping to the six-fold symmetry of the hBN crystal to write multiple zips within a region oriented integer-multiples of 60° apart. This extends the user-controlled method of introducing simple edges on the interior of existing flakes into a way to generate sharp vertices, or even to cleave customized flake shapes via unzipping. We predict the latter to be able to create polygonal flakes possessing sharp, armchair-oriented flake edges with edge lengths controllable to the sub-nm scale.

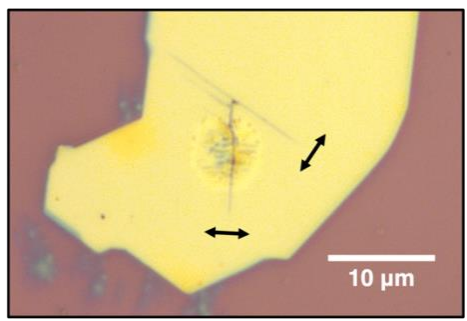

**Supplementary Figure 5.** Two zips are generated and elongated separately on a single flake with pump polarizations marked by the arrows. The individual zips, while not touching during the initial irradiation process, snap together to form 60° and 120° angles.

*SEM images of quasi-periodic gratings*

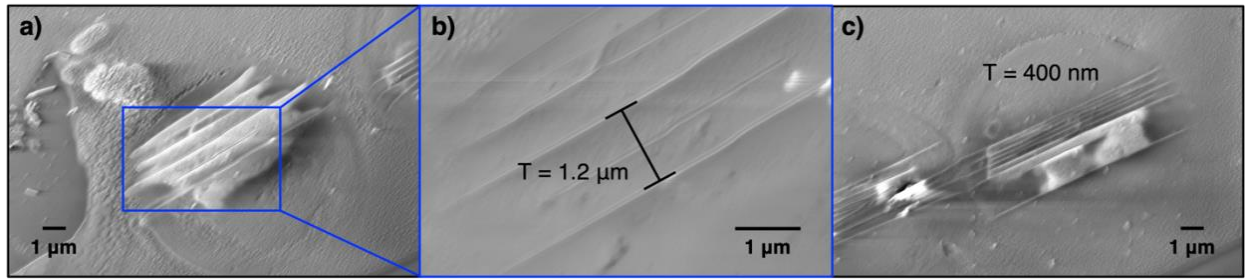

**Supplementary Figure 6.** Quasi-periodic gratings on a 38-nm-thick hBN flake on $SiO_2$/Si. The zip width here is <30 nm (below the SEM resolution in this configuration). **(a-b)** A quasi-periodic grating with a ~1.2 µm period. (The bubbles in (a) are imaging artifacts.) **(c)** Another grating on the same flake exhibiting a grating period of 400 nm. These "higher-order gratings" are less common.